\documentclass[useAMS,usenatbib,onecolumn]{mn2e}
\usepackage{psfig}
\usepackage{epsfig}

\begin{document}

\title{Influence of synchrotron self-absorption on the 21cm experiments}
\author[Zheng et al.]
  {Qian Zheng$^{1,2,3}$,
   Xiang-Ping Wu$^{1}$,
   Jun-Hua Gu$^{1}$, 
   Jingying Wang$^{4}$,
   Haiguang Xu$^{4}$
\\
  $^1$National Astronomical Observatories, 
Chinese Academy of Sciences, Beijing 100012, China; Email: zq@bao.ac.cn\\
  $^2$LERMA, Observatoire de Paris, 61, Avenue de l'Observatoire,
75014, Paris, France\\
  $^3$Graduate School of Chinese Academy of Sciences, 
Beijing 100049, China\\
  $^4$Department of Physics, Shanghai Jiao Tong University, 
800 Dongchuan Road, Shanghai 200240, China
 }

\date{Received date / accepted date}

\maketitle

\begin{abstract}
Presence of spectral curvature due to synchrotron self-absorption
of extragalactic radio sources may break down the spectral smoothness 
feature - the premise that bright radio foreground can be successfully 
removed in the 21cm experiments of searching for the epoch of 
reionization (EOR). We present a quantitative 
estimate of the effect on the measurement of the angular power 
spectrum of the low-frequency sky,  incorporating a phenomenological model, 
characterized by the fraction ($f$) of radio sources with turnover frequencies 
in 100-1000 MHz range and a broken power law for the spectral 
transition around turnover frequencies $\nu_m$, 
into the simulated radio sources over 
a small sky area of $10^{\circ}\times10^{\circ}$. 
We compare statistically the changes in their residual maps with and without
inclusion of the synchrotron self-absorption of extragalactic radio 
sources after the bright sources of $S_{150 \rm MHz}\ge100$ mJy are excised and 
the best-fitted polynomials in frequency domain on each pixel 
are further subtracted. 
It has been shown that the effect of synchrotron self-absorption
on the detection of EOR depends sensitively on the spectral 
profiles of radio sources around  the turnover frequencies $\nu_m$: 
A hard transition model described by the broken power law with 
the turnover of spectral index at $\nu_m$
would leave pronounced imprints on the residual background and therefore cause 
serious confusion with the cosmic EOR signal. However, the spectral
signatures on the angular power spectrum of 
extragalactic foreground generated by a soft transition
model, in which the rise and fall power laws of spectral distribution 
around $\nu_m$  are connected through a smooth transition spanning
$\ge200$ MHz in characteristic width, can be fitted and consequently 
subtracted by employment of polynomials to an acceptable degree 
($\delta T < 1$ mk). As this latter scenario seems to be favored by
both theoretical expectation and radio spectral observations, we conclude
that the contamination of extragalactic radio sources of synchrotron 
self-absorption in the 21cm experiments is probably very minor. 
\end{abstract}


\vskip -3in  

\section{Introduction}

While the redshifted 21cm neutral hydrogen emission is believed to 
be the most promising tool in the probe of the epoch of reionization (EOR) 
and the dark ages, the signal from the high-redshift universe is deeply 
buried under the extremely bright foreground dominated by our Galaxy 
and extragalactic sources. An unprecedented level of foreground removals 
down to 5 orders of magnitude should therefore be required to 
'excavate' the cosmic signal even in a statistical manner such as 
the employment of power spectrum  
(for recent reviews see \citealt{furlanetto06}; \citealt{morales10}; 
\citealt{pritchard11}). Yet, it is generally agreed 
among the 21cm cosmology community that the foregrounds at low frequencies 
should exhibit a featureless spectrum as a result of the spectral 
smoothness of the synchrotron and free-free emission, 
and subtracting a smooth component
characterized usually by a power-law or polynomial in frequency domain 
from the redshifted 21cm observations  
should allow one to remove the foreground contamination to a desired degree.   
Moreover, such an algorithm can be performed in either the visibility (uv) 
space  (\citealt{zaldarriaga04}; \citealt{liu09}; \citealt{harker10}; etc.) or 
the image space (\citealt{furlanetto04}; \citealt{santos05}; 
\citealt{wang06}; \citealt{dimatteo04}; etc.), 
and even directly for the angular power spectrum 
(\citealt{ghosh11a,ghosh11b}; \citealt{cho12}).  
It is, of course, essential that all bright sources in the field of view 
should be successfully excised before the
subtraction to both ensure a large dynamical range ($\sim10^5$) and 
reduce the shot noise arising from their Poisson distribution.
In other words, the foreground removals deal primarily with 
the unresolved foreground sources.  

Indeed, in our interested low frequency range, say from 50 MHz to 200 MHz, 
corresponding to 21cm neutral hydrogen line at redshifts between 6 and 27,      
a power-law or polynomial provides an accurate approximation  
of the spectral distribution for most cases of extragalactic radio sources.
However, it has been known for several decades that some of the extragalactic
radio sources including galaxies, QSOs and even clusters of galaxies 
exhibit curvatures in their spectral distributions, 
manifested by the change in their 
spectral indices approaching a constant value of $2.5$ below
the so-called cutoff or turnover frequency (e.g. \citealt{kellermann69};
 \citealt{laing80}; \citealt{andernach80}; \citealt{hodges84}).
A natural and also widely accepted explanation for the spectral curvature 
is the synchrotron 
self-absorption (\citealt{pacholczyk70}; \citealt{tsvetanov81}), though 
other mechanisms such as inverse Compton loss and free-free absorption 
or their combination may partially contribute to the phenomenon.
For any given wavebands particularly in the low frequency range, 
the fraction of the extragalactic radio sources
with spectral curvature is very uncertain. In their early work, 
\cite{kellermann69} compiled 30 radio sources that
show strong evidence of spectral curvature with turnover frequencies
from $\sim10$ MHz to $\sim1$ GHz. \cite{steppe95} observed 76
radio sources at millimeter wavelength and found about 5 sources
whose turnovers occur at frequencies below 1 GHz. Probably these 
two observations would give us a sense that the fraction of the 
radio sources with turnover frequencies below 1 GHz is not
negligibly small. 

Although a number of newly proposed algorithms for foreground removals do not
necessarily depend on the assumption of spectral smoothness of the  
foregrounds (e.g. \citealt{harker09}; \citealt{harker10b}; \citealt{chapman12}),
here we would still concentrate on the conventional method which 
makes an attempt at fitting and subtracting a polynomial of the form
$\log T_{\rm fit}=\sum_{i=0}^{N_{\rm poly}}a_i\log(\nu/\nu_0)^i$
to the observation in either image space or Fourier (uv) space,
where $N_{\rm poly}$ is the order of polynomial 
and $\nu_0$ is a characteristic frequency 
(e.g. \citealt{wang06}; Jeli\'c et al. 2006; 
 \citealt{pritchard10}). We have chosen the $\log(T)-\log(\nu)$ polynomial 
rather than the $T-\nu$ form for convenience, and the two forms of polynomial 
yield essentially similar result (\citealt{petrovic11}). 
The overall shape of the best-fit polynomial is governed by $a_1$, the 
leading component of power index, and the remaining high order terms 
for $i\ge2$ provide
only a minor correction to the deviation of power index from the dominant one
$a_1$.  With the best-fit parameter $a_1<0$ for the integrated 
foreground of extragalactic radio sources in low frequency range of
50 MHz$\le\nu\le$200 MHz, the commonly adopted form of polynomial 
describes actually a monotonically decreasing function of $\nu$.
Although it was claimed that the foreground is not perfectly flat due to
synchrotron self-absorption, it is believed that the current fitting
and subtracting techniques are still able to 
filter out the slowly varying foreground along the line of sight 
(\citealt{morales10}). 
Yet, to what extent the contamination of radio sources with spectral
curvature in the 21cm experiments can be eliminated has actually 
remained unclear so far. At least there has been no quantitative 
estimate of this particular effect in the literature. 
It seems likely that any spectral structures of foreground sources 
after performing the fitting and subtraction of a polynomial
would be treated as residuals, which could mimic or cause confusion 
with the cosmic EOR signal.  

In this paper we demonstrate quantitatively the effect of spectral 
curvature of extragalactic radio sources as a result of synchrotron 
self-absorption on the evaluation of the angular power spectrum of 
low-frequency sky. This is achieved by comparing the angular power 
spectra of the redshifted 21cm sky with and without the inclusion of
spectral curvature for extragalactic radio sources after the same method
of foreground removals is employed. Our radio foreground is constructed 
from the simulation of \cite{wilman08} with a modification of power
index for the radio sources of spectral curvature, whereas the fraction of 
the radio sources of synchrotron self-absorption in a given waveband
remains to be a free parameter. We work straightforwardly in the image 
space furnished by the numerical simulation containing no EOR signal at all,
with emphasis only on the effect of spectral curvature on foreground removals.
Throughout this paper we assume the concordance cosmological
model ($\Lambda$CDM) with the following choice of cosmic parameters: 
$\Omega_{\rm M}=0.27$, $\Omega_{\Lambda}=0.73$, 
$\Omega_bh^2=0.0224$, $h=0.71$, $n_s=0.93$ and $\sigma_8=0.84$.


\section{Extragalactic foregrounds}

There have been many theoretical explorations both analytically and 
numerically in recent years on modeling of the redshifted 
21cm foregrounds (e.g. \citealt{deoliveiracosta08}; 
\citealt{jelic08}; \citealt{bowman09}; 
\citealt{singal10}; \citealt{vernstrom11}; \citealt{liu11}).
We adopt the simulated low-frequency sky by \cite{wilman08}, in which  
the observed luminosity functions of radio sources are incorporated with 
the underlying cosmic dark matter density field out to redshift of z=20.
A total of five distinct radio source types are included in their 
semi-empirical simulation:  radio quiet AGNs, FRI and FRII AGNs, 
quiescent star-forming galaxies and star bursting galaxies. 
About 320 million sources down to 10 nJy are identified in a sky 
field of $20^{\circ}\times20^{\circ}$ at five wavebands ranging from 151 MHz
to 18 GHz.

We extract our source catalog from the $S^{3}-SEX$ database generated by 
\cite{wilman08} over a smaller sky area of $10^{\circ}\times10^{\circ}$.   
We estimate the flux of each radio source at different 
observing frequencies between 100 MHz and 200 MHz by extrapolating its 
corresponding value at 5 GHz in terms of a single power-law of $\nu^{\alpha}$
if no spectral curvature is concerned. 
Unlike \cite{wang10} who took into account the variation of power index
for each type of radio sources in the conversion of flux at different
frequencies, we simply assume a power index 
of $\alpha=-0.75$ for AGNs including FRI and FRII types
and $\alpha=-0.7$ for star forming/bursting galaxies. 
The final catalog contains 74 million sources,
and only $0.8\%$ of them have fluxes greater than 100 mJy at 150 MHz
which are named 'bright' or resolved sources in this work and will be 
excised later. In other words,  $99.2\%$ of the sources will be 
treated as the 'unresolved' foregrounds. The simulated 
intensity map of these 'unresolved' sources with fluxes fainter than 100 mJy
is displayed in Fig. 1, in which the image
has been smoothed using a Gaussian kernel of FWHM$=1'.98$,
corresponding to a radio array with a baseline of 5 km. We will adopt this 
angular resolution,  $1'.98(\nu/150 {\rm MHz})^{-1}$, 
for the simulated sky maps at other frequencies.

We have extensively searched the literature to collect as many as possible 
the radio sources of spectral curvature due to synchrotron self-absorption, 
in an attempt to find whether there are any correlations between the 
turnover frequencies $\nu_m$ and other observables such as the total
flux, the maximum flux $S_{\rm max}$ or the fluxes at particular 
frequencies. The purpose
is to insert a more realistic model of synchrotron self-absorption 
into the above simulation by allocating the proper turnover
frequencies to the sources of spectral curvature in terms of known or 
observed quantities. Unfortunately, we failed in establishing any 
statistically meaningful relationships including the expected dependence
of $S_{\rm max}$ on $\nu_m$ within the framework of synchrotron 
self-absorption under the assumption of a constant magnetic field $B$:
$S_{\rm max}\propto \nu_m^{5/2}B^{-1/2}$. 
Fig. 2 shows the peak luminosities $P_{\rm max}$, corresponding to $S_{\rm max}$ 
after the correction of cosmic expansion, against $\nu_m(1+z)$ 
for 39 sources (14 galaxies + 25 AGNs)  with observing cutoff frequencies 
below 1 GHz  compiled from three observations made 
by  \cite{kellermann69}, Hodges et al. (1984)
and \cite{steppe95}. Although $P_{\rm max}$ seems to have   
a weak increasing trend with turnover frequencies $\nu_m(1+z)$, 
the actual application of the relation in construction of the 
synchrotron self-absorption model turns to be difficult due to 
poor statistics and large scatters. For example,  a $95\%$ confidence interval
of the best-fit spectral index for the $P_{\rm max}-\nu_m$ relation is 
in the range of $0.14-1.39$.   
This probably indicates that magnetic fields demonstrate significant 
variations among radio sources. 

Keeping in mind the weak dependence, if any, of $\nu_m$ on other observables, 
we incorporate the synchrotron self-absorption feature into
the simulated radio source catalog from the $S^{3}-SEX$ database simply by 
specifying the fraction of radio sources of spectral curvatures in a 
given frequency range while 
the sources of self-absorption are chosen randomly and their turnover 
frequencies follow a uniform distribution over the waveband. 
This is consistent with the observational claim that
the turnover frequency may occur anywhere over the frequency range of
observations (e.g. \citealt{kellermann69}).  
We set a waveband of 100-1000 MHz in the restframe of the sources 
to include all the sources out to redshifts $z=4$ for our interested 
observing frequency range of 100-200 MHz. Once a source in the catalog 
is chosen to be of spectral curvature in 100-1000 MHz, 
its spectral index below  the cutoff frequency $\nu_m$ 
will be set to asymptotically approach 2.5. 
An immediate consequence of such a turnover of spectral index is that 
the radio flux of the source would drop dramatically with deceasing 
frequency below $\nu_m(1+z)$. This may explain the recent PAPER 
measurement at 145 MHz that among the $\sim500$ identified sources 
the only source, 0008-412, with spectral turnover at
$\sim500$ MHz shows no counterpart to the 408-MHz Molonglo Reference Catalog  
(\citealt{Jacobs2011}). Therefore, inclusion of synchrotron 
self-absorption alters both the overall spectral index 
and brightness of the low-frequency radio sky. 

Instead of invoking a sophisticated model for the spectral transition 
of radio source around the cutoff frequency $\nu_m$ within the
framework of synchrotron self-absorption, we assume an analytic 
profile for the sake of simplicity:  A broken power law of  $\nu^{2.5}$ and 
$\nu^{-|\alpha|}$ is used to characterize the asymptotic spectral profile 
at $\nu\ll\nu_m$ and $\nu\gg\nu_m$, respectively. 
To avoid a sudden change of spectral index across $\nu_m$, 
a soft or smooth transition around $\nu_m$ is accomplished by 
two parabola curves which match asymptotically with the 
power law at each side of the
turnover frequency $\nu_m$. The third parameter, in addition to
the peak flux  $S_{\rm max}$ and turnover frequency $\nu_m$,  
is required in such model: the characteristic frequency width $w$,  
defined  as the frequency difference between the two points 
at which the slope of each parabola is equal to that 
of the corresponding asymptotic power law.  A visual examination 
of the available spectral distributions of the radio sources of synchrotron 
self-absorption (\citealt{kellermann69}; Hodges et al. 1984
and \citealt{steppe95}) reveals that a typical value of $w$ takes  
a few hundreds of MHz. We choose an extreme value of $w=0$, called 
the hard transition model, to maximize the influence of spectral 
curvature, and a 'conservative' value of $w=$200 MHz, called 
the soft transition model, to provide a more realistic estimate 
of the effect of synchrotron self-absorption on the foreground. 
Fig. 3 shows an example of the hard transition model and of the 
soft transition model of $w=200$ MHz at  $\nu_m=200$ MHz, respectively,
in which the peak flux is assumed to be $S_{\rm max}=1$ Jy for the latter. 
Intuitively, the soft transition model seems to provide a more natural and 
reasonable description for the spectral turnover in the process of 
synchrotron self-absorption, and the discontinuously sharp peak at $\nu_m$ 
in the hard transition model turns to be difficult to handle and filter out.

Now, the last free parameter in our model of synchrotron self-absorption 
is the fraction or percentage, $f$, of radio sources with turnover
frequencies in the waveband of 100-1000 MHz. Because very little 
is known about the constraint on $f$ in terms of current radio observations, 
we work with three choices of $f$, covering the 'extreme' value 
of $10\%$, the moderate case of $1\%$, and 
the lower, probably more realistic value of $0.1\%$.  
Finally, the case of no spectral curvature ($f=0\%$) 
serves as our fiducial model for comparison. Fig. 4 shows the
average surface brightness temperature of our simulated maps 
of $10^{\circ}\times10^{\circ}$ versus frequencies for the four choices 
of $f$, together with the residual 
after the best-fitted polynomial of $N_{\rm poly}=3$  
over the whole frequency range from 100 MHz to 200 MHz is subtracted. 
We demonstrate the results for the hard transition model only, which
gives rise to a maximum estimate of the effect of spectral curvature 
on the global 21cm sky intensity. It appears that 
except for the slightly lower amplitudes for $f>0$ as a result of 
synchrotron self-absorption, the overall 
sky temperatures show no significant difference in shape as compared with
the result of our fiducial model of $f=0$. This allows us to
subtract the foregrounds with and without inclusion of synchrotron 
self-absorption to essentially the same level as shown in Fig.4, 
in which the residuals reflect actually the current accuracy level 
of our foreground removals based on the polynomial fitting method.  
Indeed, the effect of spectral curvature has become insignificant by  
ensemble-averaging of surface brightness temperature over the sky. 
Consequently,  the experiments of searching for the global EOR signatures 
such as EDGES (\citealt{bowman07}; \citealt{bowman10}) may be 
unaffected by the presence of foreground radio sources of
synchrotron self-absorption.

\section{Angular power spectrum}

\subsection{Foreground removals and shot noise}

As our focus is not on the study of foreground removal techniques, we 
apply directly the pixel-by-pixel algorithm of \cite{wang06} for
subtraction of the simulated extragalactic sources.
Before we perform the foreground removal, there are two critical
assumptions that we have to make. First, all the bright
sources of flux greater than 100 mJy have been successfully excised.
Second, thermal noise on each pixel has already been reduced to an acceptable 
degree. Yet, the first assumption or operation by no means implies that 
the shot noise due to the Poisson distribution of extragalactic 
sources has been reduced to the desired level for the detection of 
EOR signal. Rather,  we will actually deal with the shot noise produced by 
the 'unresolved' extragalactic foregrounds.  As for the second condition
the thermal noise can in principle be reduced
by both accumulating observations and/or employing statistical methods such as 
power spectrum, provided that the noises on different pixels are independent. 

To guarantee a sufficiently large dynamical range of the foreground removal
algorithm, we adopt a short frequency range of 20 MHz rather than the 
full frequency width of 100 MHz in the fitting of a log-log polynomial. 
Consequently, a set of five log-log polynomials, with a bandwidth of 
20 MHz for each, are separately fitted to and then subtracted from 
the intensity on each pixel of the simulated backgrounds. We fix a 
$1024^2$ grid to sample the simulated image with a size of 
$10^{\circ}\times10^{\circ}$, which gives rise to an angular resolution of 
$0.^{''}6$. Here we have included no any instrumentational responses such as 
frequency-dependent field-of-view and angular resolution to lay stress only 
on the effect of the spectral curvature of extragalactic radio sources.

Even without inclusion of the noises from telescope system and the Milky
Way,  the angular power spectrum of the radio sky at a given low-frequency 
is still dominated by the shot noise of the 'unresolved' extragalactic 
sources, as demonstrated in Fig. 5 for four frequency channels. 
Such Poisson noise 
is fully characterized by the differential source counts $dN/dS$ below 
the flux cut $S_{\rm cut}$ (see \citealt{knox01}; \citealt{dimatteo02}):  
$C_{\ell}^{\rm shot}=\int_{0}^{S_{\rm cut}}S^{2}(dN/dS)dS$. 
Given the fact that the cosmological signal from EOR has a brightness 
temperature fluctuation of a few to ten mK (e.g. Zaldarriaga et al. 2004), 
the shot noise power spectrum of the foregrounds consisting of  
the unresolved' sources below  $S_{\rm cut}$ should be further 
and blind suppressed by at least two orders of magnitude in order to 
extract statistically the EOR information. This is accomplished by
various foreground removal techniques including the polynomial 
fitting algorithm. It should be pointed out that the angular power
spectrum constructed from our simulated sky is restricted within a narrow
multipole range roughly from $\ell\approx20$ to $\ell\approx1000$ due
to the small field of $10^{\circ}\times10^{\circ}$ and the smoothing 
window of FWHM$=1'.98(\nu/150 {\rm MHz})^{-1}$. The former 
yields a large cosmic variance at smaller $\ell$, while the latter 
smears out the spatial structures at larger $\ell$. We will not account
for these measurement errors in our calculation of power spectrum
because they do not alter our conclusions to be drawn below.

\subsection{Comparisons of angular power spectra: hard transition model}

We now work with the residual map at each frequency after the
best-fitted polynomials of $N_{\rm poly}=3$ are subtracted on each pixel. 
We first demonstrate the results for the  
hard transition model of spectral curvature. 
Fig. 6 displays the average surface brightness temperature 
of the residual maps for four choices of $f$, the fraction of
radio sources of turnover frequencies in the 100-1000 MHz range.
The difference between the residuals in Fig.4 and 
Fig.6 is as follows:  
the former is obtained by performing the polynomial fitting 
and subtraction on the global surface temperature after summing over the 
contribution of each pixel, whereas for the latter the polynomial fitting 
and subtraction are made on each pixel before the averaging.  
Regardless of the regular oscillation patterns in Fig.6 
which are the consequences of the polynomial subtraction, 
the overall residuals of our
fiducial model ($f=0\%$) without synchrotron self-absorption are well below
0.1 mK in the entire frequency range of 100-200 MHz, indicating the success
of the polynomial fitting and subtraction techniques in the removal of
extragalactic radio sources. However, the average residual temperature 
rises with increase of the fraction parameter $f$, 
and some of the oscillation peaks 
exceed 1 mK in the case of $f\ge1\%$. Because the average residual temperature 
at each frequency is somehow equivalent to the normalization factor 
of the corresponding angular power spectrum to be constructed, 
the large amplitude of the residual temperature of up to $\sim1$ mK 
due to the presence of synchrotron self-absorption 
may challenge the statistical extraction of the EOR signal which is expected  
to have a maximum spatial variance of 1-10 mK.  Furthermore,
it is anticipated that the amplitudes of the angular power spectra 
at different frequencies may demonstrate significant variations as a 
result of the dramatic rise and fall of their normalizations as shown in 
Fig.6.  

The angular power spectrum of the residual temperature map after subtraction 
of the best-fitted polynomials can be obtained straightforwardly 
at each frequency channel for the hard transition mode of $w=0$.
We plot in Fig. 7 the corresponding results,  
represented by $\delta T=[\ell(2\ell+1)C_{\ell}/4\pi]^{1/2}$, for the four
choices of $f$ and frequency parameters used in Fig. 4.  
While the overall power spectra maintain roughly the same shape for
the four models of $f$ and at different frequencies, the amplitudes 
of the power spectra for all the three choices of $f\ge0.1\%$ greatly 
exceed those from our fiducial model of $f=0$. In most cases the
amplitude differences can span three orders of magnitude, accounting for
a power of up to a few - tens mK for $20<\ell<1000$ on the power spectrum 
of the radio sky. 
In other words, the shot noises from the foreground residuals are 
comparable to and thus capable of confusing the cosmic EOR signal, if the 
fraction of the extragalactic radio sources of spectral curvature 
in $100-1000$ MHz is not less than $0.1\%$. Indeed, the presence of 
synchrotron self-absorption described by the hard transition model  
has already broken the spectral smoothness feature  
of extragalactic radio sources - a key assumption to beat down 
the extremely bright foregrounds in the 21cm experiments. 
It is also important to note that both the oscillation patterns 
along frequency direction (Fig.6) and
the angular power spectra of the unresolved radio foregrounds 
in the existence of synchrotron self-absorption 
look very much similar to the EOR signatures, making the two 
phenomena practically indistinguishable.  

Probably, an intuitive way to reduce the impact of 
extragalactic sources of spectral curvature in the framework of the hard 
transition model on the radio foregrounds  
is to adopt a lower flux threshold of $S_{\rm cut}<100$ mJy. Namely, 
one can identify and excise much fainter extragalactic radio sources 
to suppress further the flux level of the 'unresolved' foreground. 
Towards this end, we set $S_{\rm cut}$ down to an extremely low flux of 
$0.1$ mJy at 150 MHz
and assume that all bright sources above  $0.1$ mJy can be perfectly 
removed with existing and developing observational techniques such as 
CLEAN and its variants (e.g. \citealt{schwab84}; \citealt{cotton08}),   
peeling (\citealt{noordam04}; \citealt{vandertol07}; \citealt{intema09}), 
algebraic forward modeling (\citealt{bernardi11}), etc.. We have applied 
this fainter flux limit to the same source catalog complied above and 
repeated all the computations.  It turns out that the overall amplitudes 
of the resultant average surface brightness temperature and angular 
power spectra of the residual maps indeed decrease moderately 
as compared with the corresponding quantities in Figs 6 and 7.
However, some part of the angular power spectra 
can still maintain high powers of up to a few mK especially at large $\ell$ 
in the case of $f>0.1\%$,  indicating the inefficiency 
of eliminating entirely the effect of spectral curvature on the foreground 
removals even down to the level of $S_{\rm cut}=0.1$ mJy.  
At such a lower flux threshold, the amplitude of the angular 
power spectrum for $f=0$ is mainly governed by the noise resulting from 
polynomial fitting and subtraction, whereas the failure of removing
the spectral oscillation structures with polynomial fitting for 
the radio sources of synchrotron self-absorption 
accounts for the relatively large amplitudes of their power spectra. 
Of course, one may adopt even smaller values of 
$S_{\rm cut}$ to continue the above procedure until the amplitude of the angular 
power spectrum is made below 1 mK, if the feasibility of radio observations 
and subtraction of very faint point sources are left aside. 
Note that the average flux and angular 
power spectrum of the radio sky, rather than their residuals, vary as  
$\langle S\rangle\propto S_{\rm cut}^{2-|\alpha|}$ and 
$C_{\ell}^{\rm shot}\propto S_{\rm cut}^{3-|\alpha|}$, respectively. At the 
faint flux end, the spectral index $\alpha$ is approximately $-1$. 
Consequently, one is in principle able to suppress  both global 
flux and spatial structures of the foregrounds  by
imposing a lower flux cutoff on extragalactic sources.  
Yet, an extremely or even unreasonably lower flux threshold $S_{\rm cut}$ 
would be required to bring the foreground down to the level of 
cosmic EOR signal. 

\subsection{Comparisons of angular power spectra: soft transition model}

It is very likely that the influence of synchrotron self-absorption of 
foreground radio sources on the 21cm experiments has been exaggerated 
by the oversimplified, hard transition model for spectral turnover, 
in which the discontinuously sharp peak at $\nu_m$ can hardly be filtered
out through subtracting a set of polynomials. In other words,
the excess power in the angular power spectrum of radio foreground 
in the case of $f\neq0$ shown in Figs 6 and 7 could be an artifact of the
spectral transition model. It is essential to use the soft transition
model to clarify the issue. Indeed, in our soft transition
model the smooth transition region around turnover frequency $\nu_m$  
spans $w=200$ MHz in frequency (see Fig.3), which is  
wider than the bandwidth of 20 MHz
used in the polynomial fitting.  Such a smooth component around  $\nu_m$  
should therefore be characterized by the polynomials 
and subtracted subsequently, if the overlapping parts of the rise and fall 
power laws around  $\nu_m$ from different sources of spectral curvature 
are not serious.

Following the same procedure as for the hard transition model above, 
we display in Figs 8 and 9 the average surface brightness 
temperature and angular power spectra of the residual surface temperature 
maps for the soft transition model of $w=200$ MHz with  
$S_{\rm cut}=100$ mJy at 150 MHz.  
It appears that both quantities are roughly one order of magnitude 
smaller than the ones for the hard transition model, bringing the amplitudes
of angular power spectra down to $\delta T\la1$ mK 
for the three choices of $f\ne0$. Only in the case of low frequency 
$\nu=110$ MHz  can a slightly higher 
power of $\delta T\sim1$ mK be seen for $f=10\%$ and $\ell\ge600$.  
This indicates that the influence of spectral curvature of radio sources on 
foreground removals is only moderate for $f\ge10\%$ and at low frequencies 
$\nu\la110$ MHz, and becomes actually minor for frequencies 
beyond $\nu\ga125$ MHz.   Most likely, the characteristic width $w$ 
would not be smaller than 200 MHz in terms of available spectral observations
of radio galaxies and the fraction of radio sources of spectral curvature 
in the 100-1000 MHz range would not exceed $f=10\%$ as indicated by
existing samples such as \citealt{kellermann69} and \citealt{steppe95}. 
Therefore, the actual angular power spectra of foreground residual maps
probably have their amplitudes well below 1 mK if a large value of 
$w$ and a smaller value of $f$ parameters are taken. 

Such an apparent contradiction between the results of the hard transition model 
and the ones of the soft transition model with $w\ge200$ MHz emphasizes
the importance of understanding
and properly modeling the spectral properties of radio sources 
in the 21cm experiments. Any sharp, though weak, variations in
spectral distributions of foreground radio sources over  
a narrow band might leave observing imprints on the foreground residuals, 
and thus contaminate the detection of 21cm EOR signal. Recall that
the foreground should be suppressed down to 5 orders of magnitude 
in the 21cm experiments. 
In terms of our current understanding of synchrotron self-absorption of 
radio galaxies and AGNs both theoretically
and observationally, the spectral transition around turnover frequency 
should be a slowly varying physical process spanning typically 
a few hundreds of MHz in frequency. It seems that the soft transition 
model of $w\ge200$ MHz provides a more natural and reasonable description 
of the spectral turnover due to synchrotron self-absorption, 
despite that it is only a phenomenological model.

\subsection{Comparisons of angular power spectra: multi-frequencies}

We can probably go one step further in exploring the effect of spectral 
curvature of radio sources 
on the multi-frequency angular power spectrum of the foregrounds.
This is motivated by the widely accepted hypothesis that the cosmic 21cm
signals measured at two frequencies $\nu$ and $\nu+\Delta\nu$ are 
uncorrelated if their frequency difference $\Delta\nu$ exceeds typically 
$\sim1$ MHz, corresponding to a comoving scale of $\sim20$ Mpc at $z=10$,  
while the foregrounds beyond $\Delta\nu>1$ MHz should still remain 
strongly correlated as a result of continuum radiation 
(e.g. \citealt{zaldarriaga04}; \citealt{santos05}; \citealt{wang06}; 
\citealt{datta07}). Therefore, it is possible to extract the weak cosmic 
21cm signal in the presence of strong foregrounds
by fitting a polynomial to the multi-frequency angular power spectra 
$C_{\ell}(\nu,\nu+\Delta\nu)$ only for large intervals of $\Delta\nu$ 
in frequency domain and then subtracting the best-fit result from 
the observed $C_{\ell}(\nu,\nu+\Delta\nu)$.
Application of this technique to the GMRT observation has been made recently
by \citealt{ghosh11a} and \citealt{ghosh11b}.  Equally, One can    
work with the three-dimensional power spectrum by removing the 
foreground contamination in terms of the power at lower radial component 
of wavenumber $k_{||}$.

Following the conventional definition, we calculate the multi-frequency 
angular power spectrum through 
$C_{\ell}(\Delta\nu)=C_{\ell}(\nu,\nu+\Delta\nu)=
\langle a_{\ell m}(\nu)a^*_{\ell m}(\nu+\Delta\nu)\rangle$, 
where $a_{\ell m}(\nu)$ is the coefficient of spherical harmonics at frequency
$\nu$, and $\langle \cdot\cdot\cdot\rangle$ denotes the ensemble average. 
We take the soft transition model for description of synchrotron 
turnover and the same four choices of $f$ as the above for the fraction 
of radio sources of spectral curvature. In Fig.10(a) we illustrate
the multi-frequency angular power spectra $C_{\ell}(\Delta\nu)$ at 
$\ell=400$ and $\ell=1200$ derived 
from our simulated maps with a frequency resolution of $0.1$ MHz for
$\Delta\nu$ ranging from 0 to 5 MHz.  We first fit a polynomial of
$N_{\rm poly}=3$ to $C_{\ell}(\Delta\nu)$ over the entire frequency interval
out to $\Delta\nu=5$ MHz, and then subtract the best-fit polynomial from
$\Delta\nu$ to get the corresponding residual [see Fig.10(b)] for each 
choice of $f$ parameter. 
In all the circumstances,  the residuals appear to be well below 0.1 mK, 
indicating that the foreground contaminations have been perfectly removed. 
Now we confine the polynomial fitting to the data 
in $\Delta\nu=0.5$-$5$ MHz. Namely, the data within $\Delta\nu=0.5$ MHz are 
excluded in the fitting in orer to concentrate only on the foreground 
contribution. The best-fit polynomials are finally subtracted from the 
total $C_{\ell}(\Delta\nu)$ over the entire range of  $\Delta\nu$.  
It turns out that all the residuals including those in the lower frequency
intervals of $\Delta\nu\le0.5$ MHz are smaller than 0.1 mK. Therefore,
we conclude from these simple exercises that the synchrotron 
self-absorption of foreground radio sources has no significant influence 
on the construction of cosmic 21cm signal using 
the multi-frequency angular power spectrum technique.

\section{Discussion and conclusions}

Instead of invoking a sophisticated description for the synchrotron 
self-absorption of extragalactic radio sources, we have taken a 
model characterized by three parameters to study 
the possible contamination of the foreground radio sources of 
spectral curvature in the upcoming 21cm experiments dedicated 
to the detection of EOR signatures:  the fraction ($f$)
of extragalactic sources whose spectral curvature occur in $100-1000$ MHz
in the restframe of the sources, the turnover frequency ($\nu_m$)
and the frequency width $w$ that specifies the spectral transition region
around $\nu_m$.  To highlight the effect of the 
spectral curvature, we have included neither the Milky Way foreground 
nor the 21 emission background of neutral hydrogen from EOR in our
computations. We have used the simulated sky map of  
$10^{\circ}\times10^{\circ}$
to calculate and compare the angular power spectra with and without
inclusion of the synchrotron self-absorption of extragalactic radio 
sources. We have then made a thorough comparison of the results 
between the two cases after subtracting the best-fit
log-log polynomials of order $N_{\rm poly}=3$ on each pixel. 

Presence of the synchrotron self-absorption of extragalactic radio sources
breaks down the monotonically increasing/decreasing spectral feature, 
which may post a challenge to   
the premise that the strong foreground can be perfectly removed 
in the 21cm experiments. It is therefore essential that the fraction, 
the turnover frequencies and the spectral transition profile  
of radio sources be robustly determined 
so that their influence on the 21cm experiments can be 
evaluated quantitatively and precisely. While the synchrotron 
self-absorption has been
unambiguously observed in some of the radio sources especially in 
low frequencies ranging from 50 to 200 MHz, the above three key 
parameters/properties and their cosmic evolution have remained 
somehow uncertain so far. 
Our simple phenomenological model, in which the fraction ($f$) 
of the extragalactic radio sources of spectral curvature is allowed to vary, 
provides the first estimate of how and to what extent the 
synchrotron self-absorption of foreground radio sources may contaminate 
the measurement of the EOR signal in the 21cm experiments. 

It has been shown that the effect of synchrotron self-absorption of
extragalactic radio sources on the foreground removals in the 21cm 
experiments depends sensitively on the spectral distribution around the
turnover frequency $\nu_m$: a hard transition at $\nu_m$ leads to a 
sudden change of spectral index, which cannot be  
fitted and subtracted by the slowly varying form of polynomials. 
Consequently, if extragalactic radio sources whose turnover frequencies occur 
in the $100-1000$ MHz waveband are not less than $0.1\%$ of the total 
radio sources, the residuals of the unresolved radio foreground, manifested 
by either global flux or angular power spectra, would exceed the background 
EOR signal, increasing the difficulties of separating the cosmic EOR 
signatures from spectral structures of extragalactic radio sources. 
This is because 
the existence of synchrotron self-absorption characterized by the hard
transition model has altered the underlying 
assumption about the intrinsic spectral smoothness of radio foreground 
along frequency space. 
While the hard transition model for the spectral profile of 
synchrotron self-absorption exaggerates  
the contamination of the foreground radio sources in the measurement 
of 21cm cosmic background, it demonstrates how foreground removals
in the 21cm experiments are affected by sharp variations in the spectral 
profiles of radio sources even if the fraction of such radio sources 
is relatively small.

Theoretically, it is expected that the synchrotron self-absorption 
of cosmic radio sources should exhibit a smooth, rather than hard, 
spectral transition around turnover (e.g. \citealt{odell70}).  
Such a scenario is further supported by a number of radio 
spectral observations of QSOs, galaxies and even 
clusters of galaxies (e.g. \citealt{kellermann69};
 \citealt{laing80}; \citealt{andernach80}; Hodges et al. 1984).
This may alleviate the concern raised in the hard transition model that 
the effect of synchrotron self-absorption may leave pronounced 
imprints on the 21cm background. Indeed, using the soft transition
model of $w=200$ MHz
for the radio sources of synchrotron self-absorption, we have
successfully reduced the contribution of unresolved extragalactic sources 
to an acceptable degree ($\delta T <1$ mk)
because the smooth spectral transition 
spanning a few hundreds of MHz around turnover frequencies 
can be nicely fitted and subsequently subtracted by a set of 
polynomials over a frequency width of 20 MHz for each.

The presence of spectral curvature in extragalactic radio sources
does increase the technical difficulties 
of subtracting the foreground point sources and identifying the EOR 
signatures. However, if the transition of spectral index 
for extragalactic radio sources proceeds rather smoothly 
around turnover frequencies as indicated presently 
by both physical and observational 
scenario, the synchrotron self-absorption of extragalactic radio sources 
should not constitute a major obstacle to the foreground removals
for the extraction of EOR signal in the 21cm experiments. 
Finally, it should be noted that 
our calculations and results are not confined to the radio sources 
with synchrotron self-absorption alone but applicable to 
any types of radio sources of spectral curvatures,
because there is no particular physics behind our phenomenological model.

\section{Acknowledgments}
This work was supported by the Ministry of Science and 
Technology of China under Grant 2009CB824900, and  
the National Science Foundation of China under Grant 
10878001/10973010/11125313. The authors would like to 
acknowledge an anonymous referee for constructive suggestions
and insightful comments.


\clearpage
\begin{figure}
\begin{center}
\psfig{file=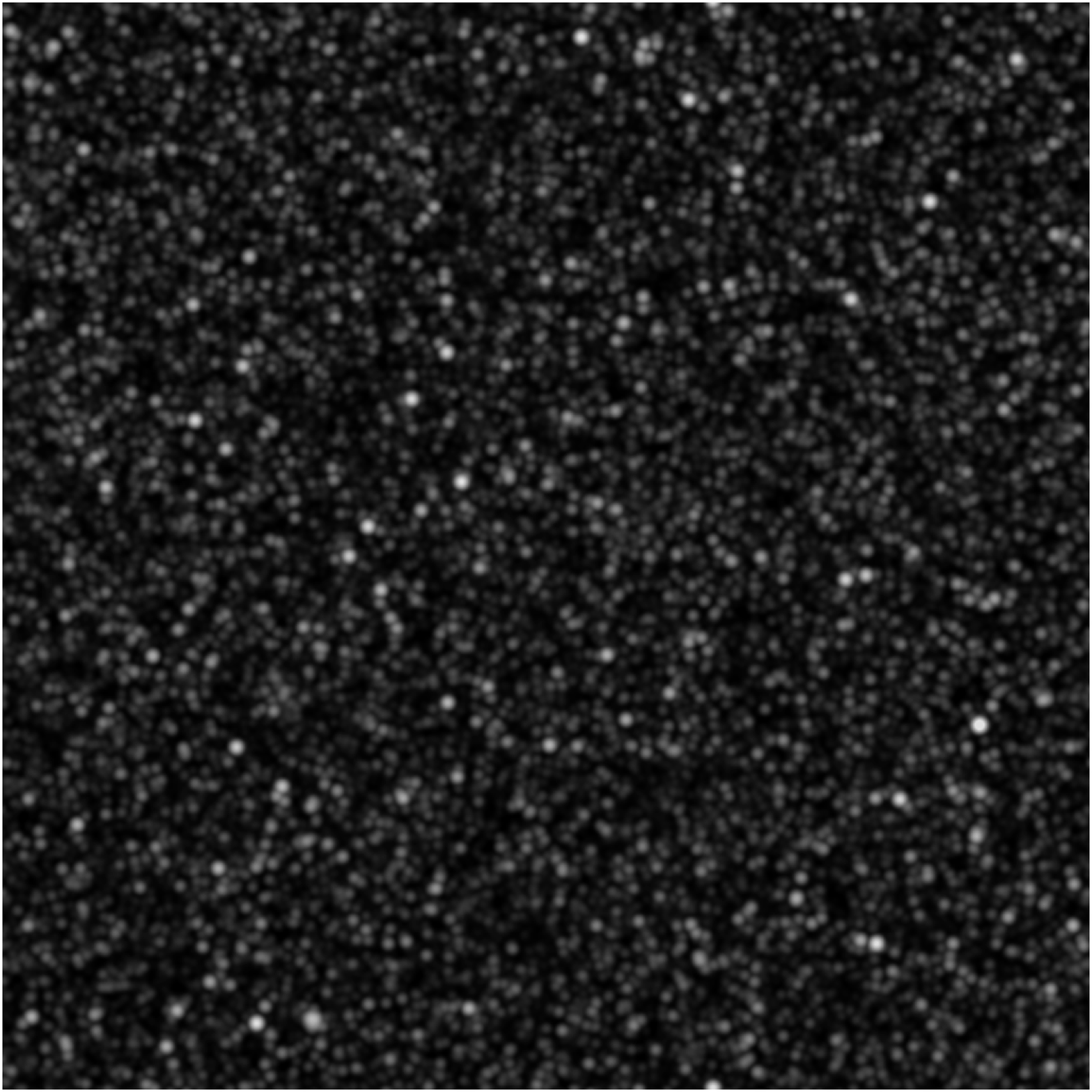,width=12.0cm}
\end{center}
\caption{Simulated sky map of the 'unresolved' radio sources below 
100 mJy at frequency $\nu=150$ MHz. The map has an 
angular size of $10^{\circ}\times10^{\circ}$ and is smoothed with a 
Gaussian kernel of FWHM$=1'.98$.}
\end{figure}

\begin{figure}
\begin{center}
\psfig{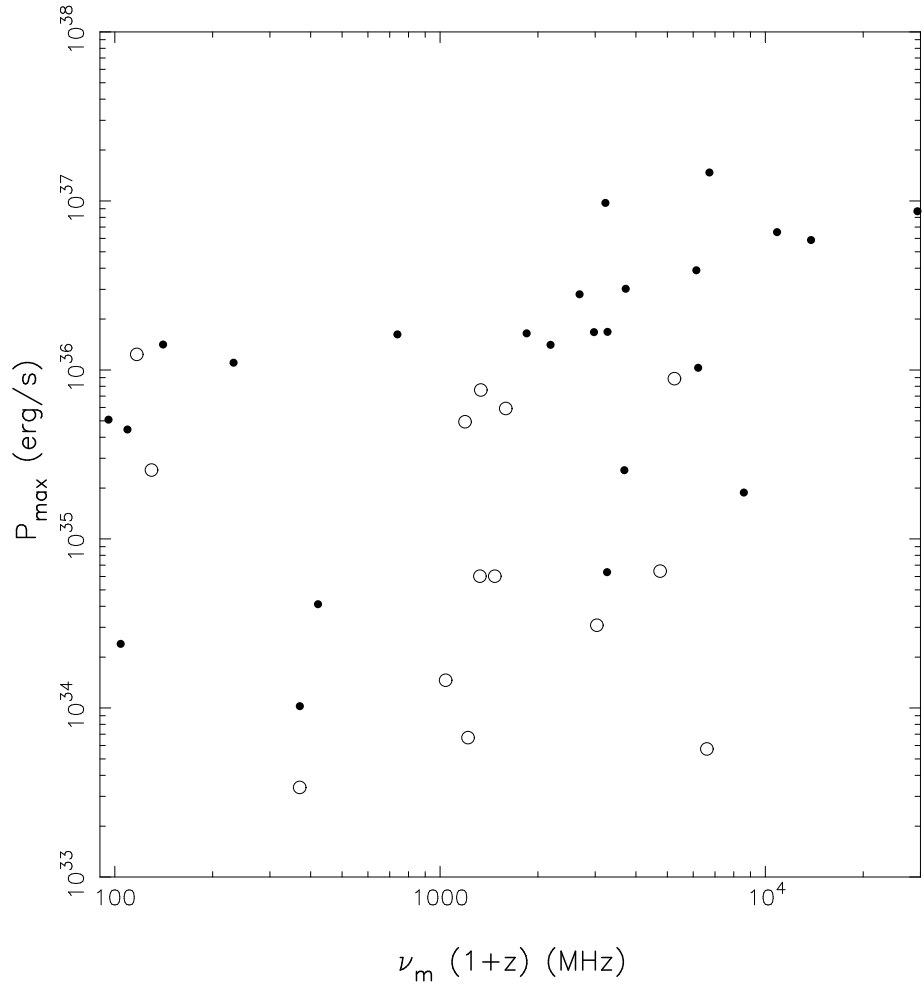}
\end{center}
\caption{Peak radio luminosities $P_{\rm max}$ are plotted against 
turnover frequencies $\nu_m(1+z)$ for 39 radio sources in which 
the spectral curvatures are successfully interpreted as being due to 
synchrotron self-absorption. Data are compiled from 
Kellermann \& Pauliny-Toth (1969), Hodges et al. (1984) 
and Steppe et al. (1995). 
Galaxies and AGNs are denoted by open and filled circles, respectively.}
\end{figure}

\begin{figure}
\begin{center}
\psfig{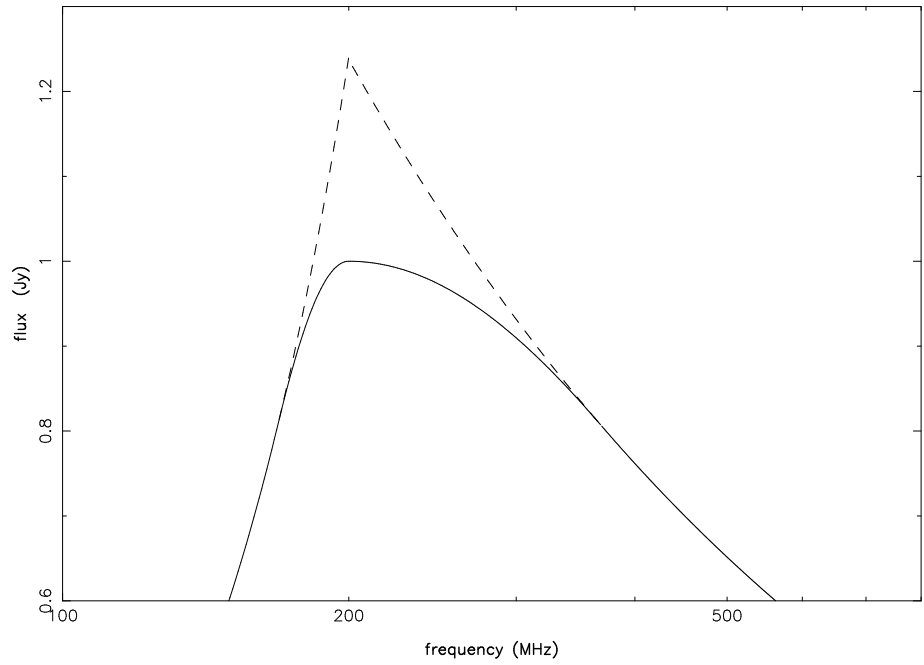}
\end{center}
\caption{An example of the hard transition model (dashed line)  
and the soft transition model of $w=200$ MHz (solid line)
at $\nu_m=200$ MHz.} 
\end{figure}

\begin{figure}
\begin{center}
\psfig{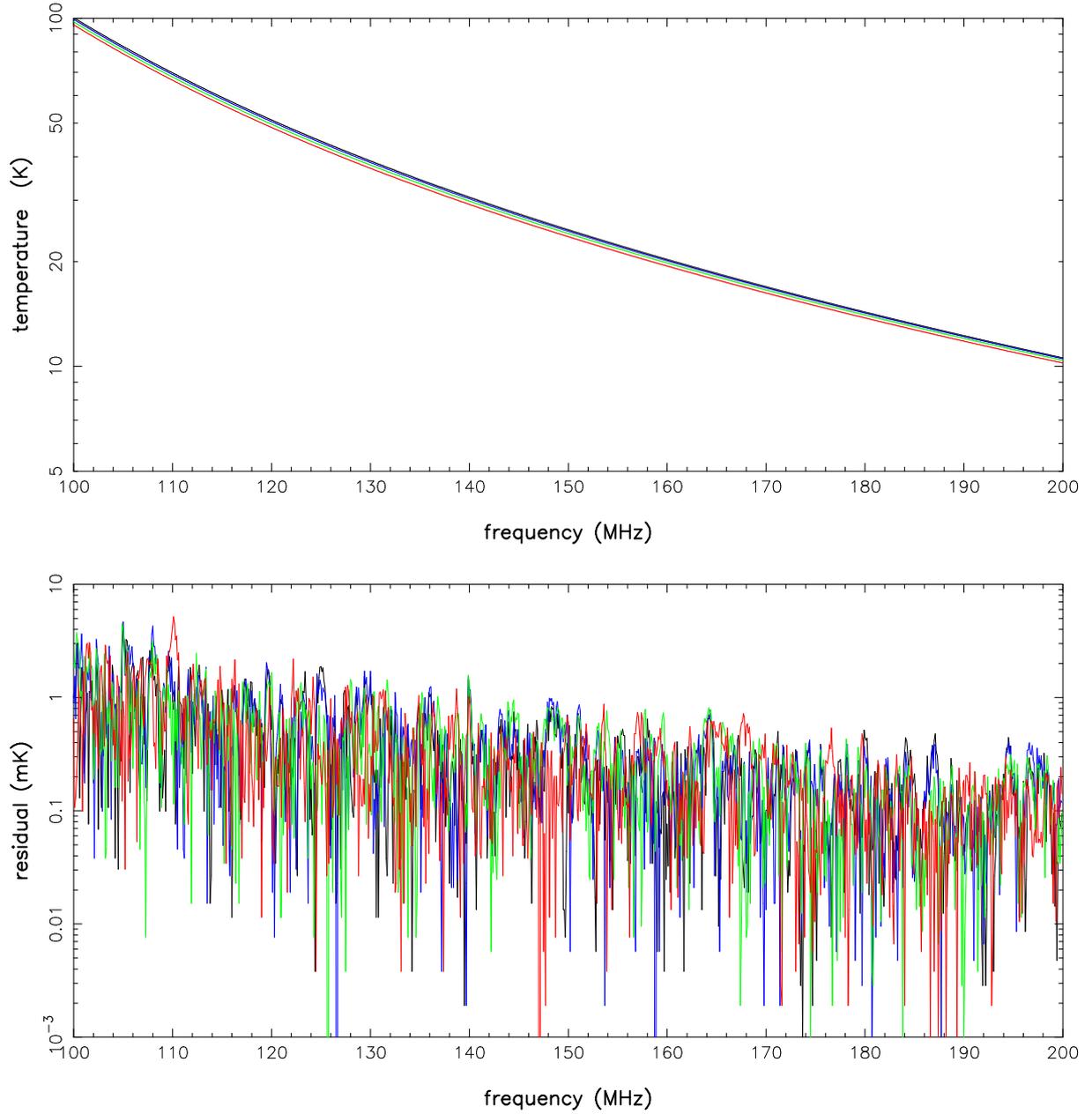}
\end{center}
\caption{Top panel: Average surface brightness temperature 
versus frequency for four choices of $f$, the fraction of radio sources 
with turnover frequencies in 100-1000 MHz range: 
$f=10\%$ (red line), 
$f=1\%$  (green line),
$f=0.1\%$  (blue line),
and $f=0\%$  (black line).
Bottom panel: The corresponding residuals for the four models of $f$   
after the best-fitted polynomials of $N_{\rm poly}=3$ are subtracted. }
\end{figure}

\begin{figure}
\begin{center}
\psfig{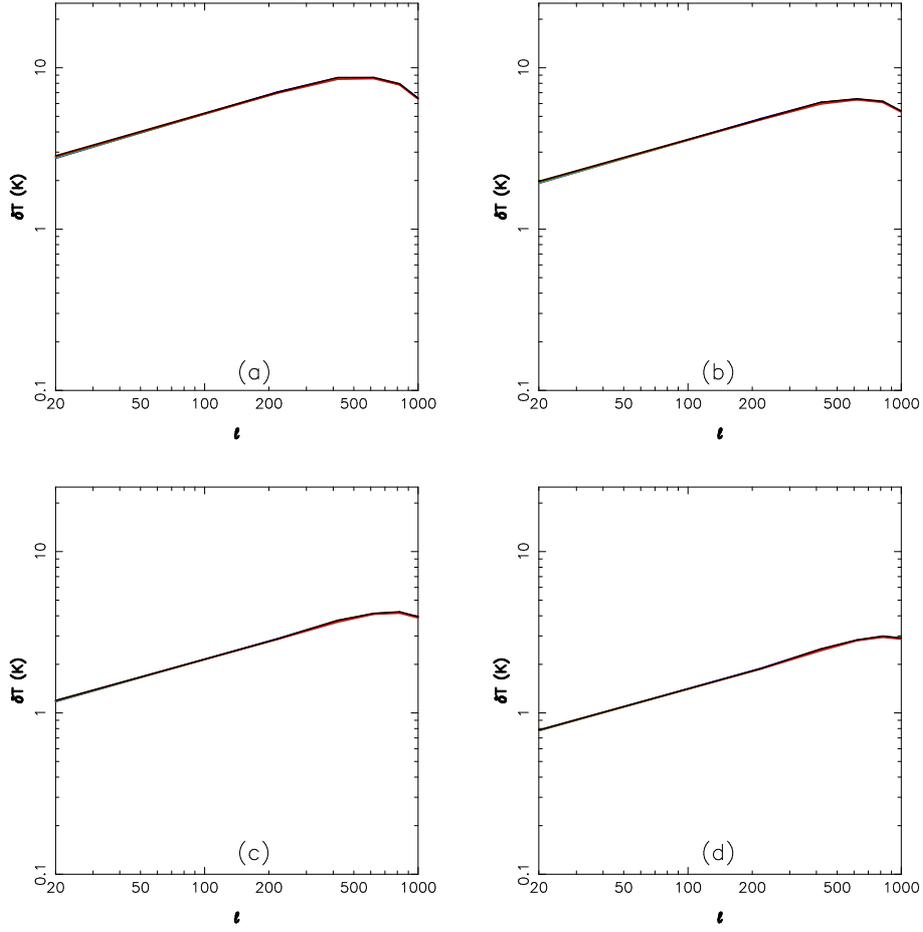}
\end{center}
\caption{Angular power spectra of the 'unresolved' foregrounds down 
to $S_{150\rm MHz}=100$ mJy for the four models of $f$ and at four 
frequencies: (a)$\nu=110$ MHz, (b)$\nu=125$ MHz,
(c)$\nu=150$ MHz and (d)$\nu=175$ MHz. The power spectra, represented by 
$\delta T=[\ell(2\ell+1)C_{\ell}/4\pi]^{1/2}$, are 
dominated by the shot noises of the radio sources and rather insensitive 
to the choice of $f$.}
\end{figure}

\begin{figure}
\begin{center}
\psfig{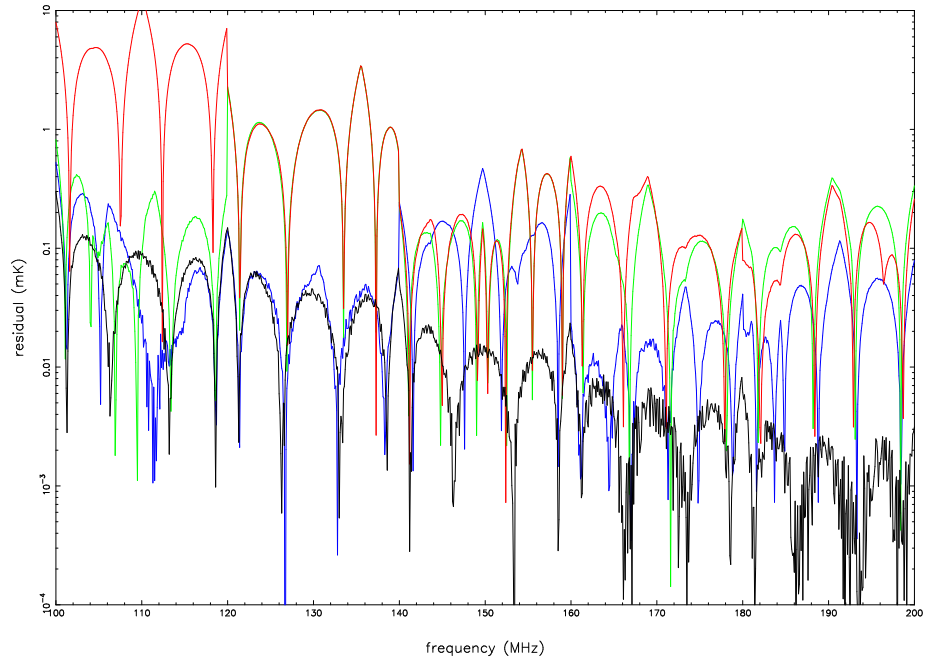}
\end{center}
\caption{Average surface brightness temperature of the residual maps after the 
best-fit polynomials over a bandwidth of 20 MHz are subtracted 
on each pixel for the hard transition model. 
The same notations as in Fig.4 are adopted.}
\end{figure}

\begin{figure}
\begin{center}
\psfig{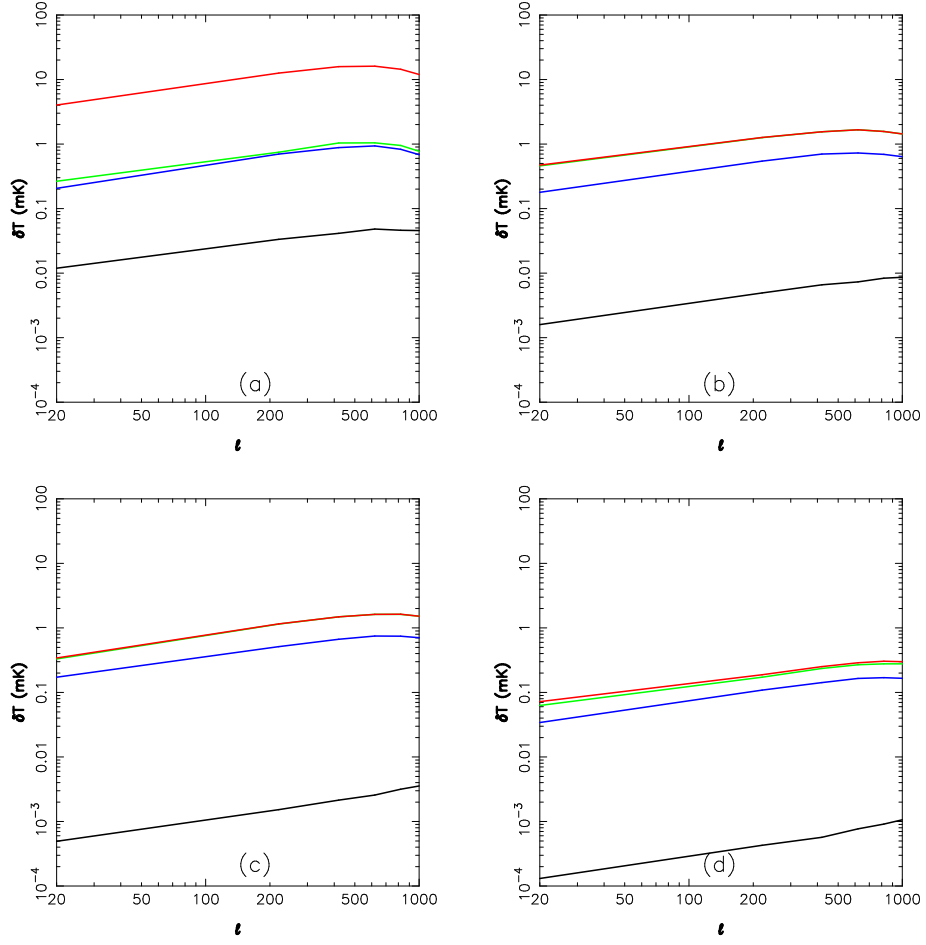}
\end{center}
\caption{Angular power spectra of the residual maps, corresponding 
to the four plots in Fig.5, after subtracting
the best-fitted polynomials on each pixel. The same notations 
as in Fig. 4 are shown for the four models of $f$.}
\end{figure}

\begin{figure}
\begin{center}
\psfig{file=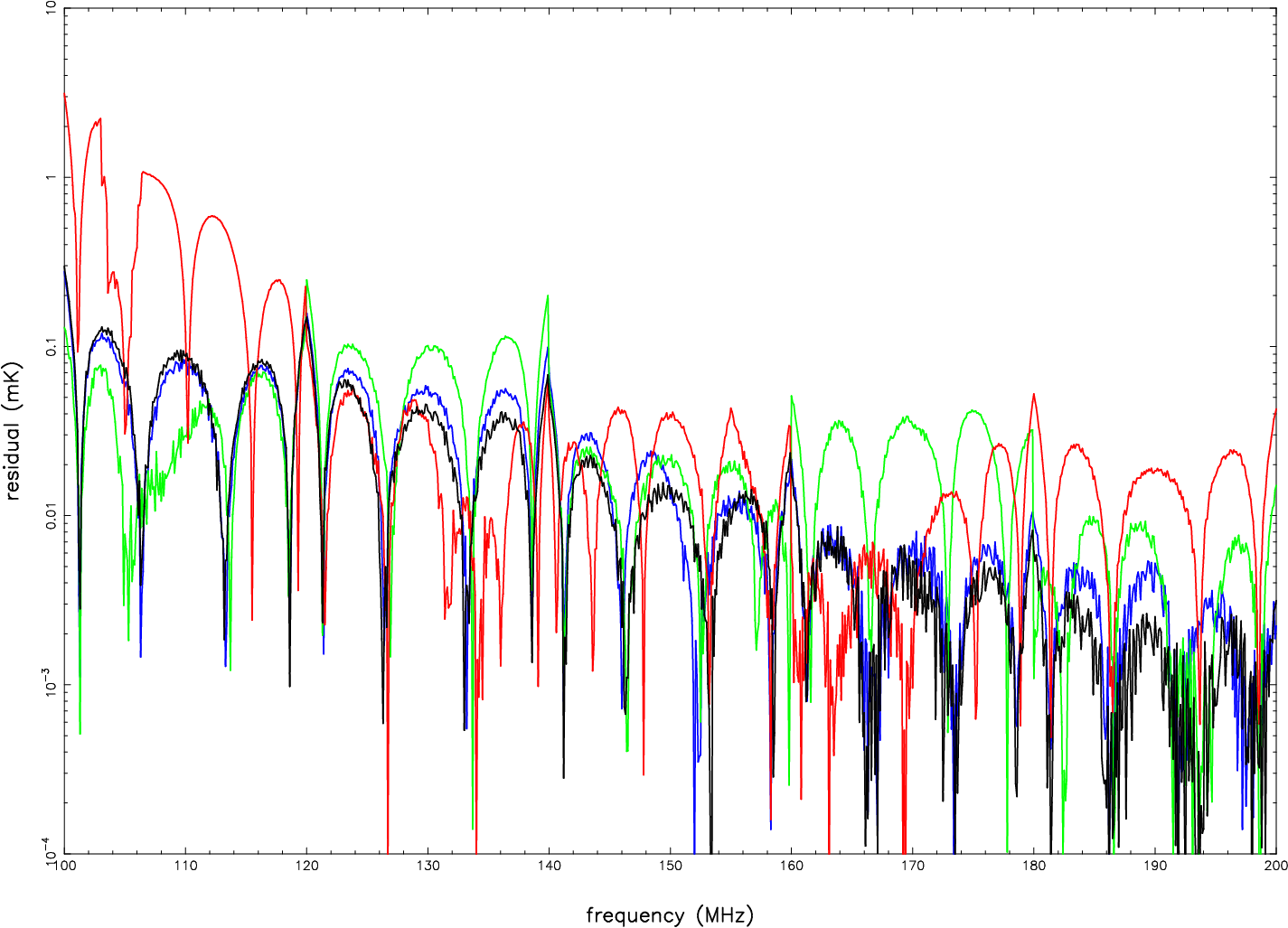,angle=-90,width=12.0cm}
\end{center}
\caption{The same as Fig. 6 but for the soft transition model of $w=200$ MHz.}
\end{figure}

\begin{figure}
\begin{center}
\psfig{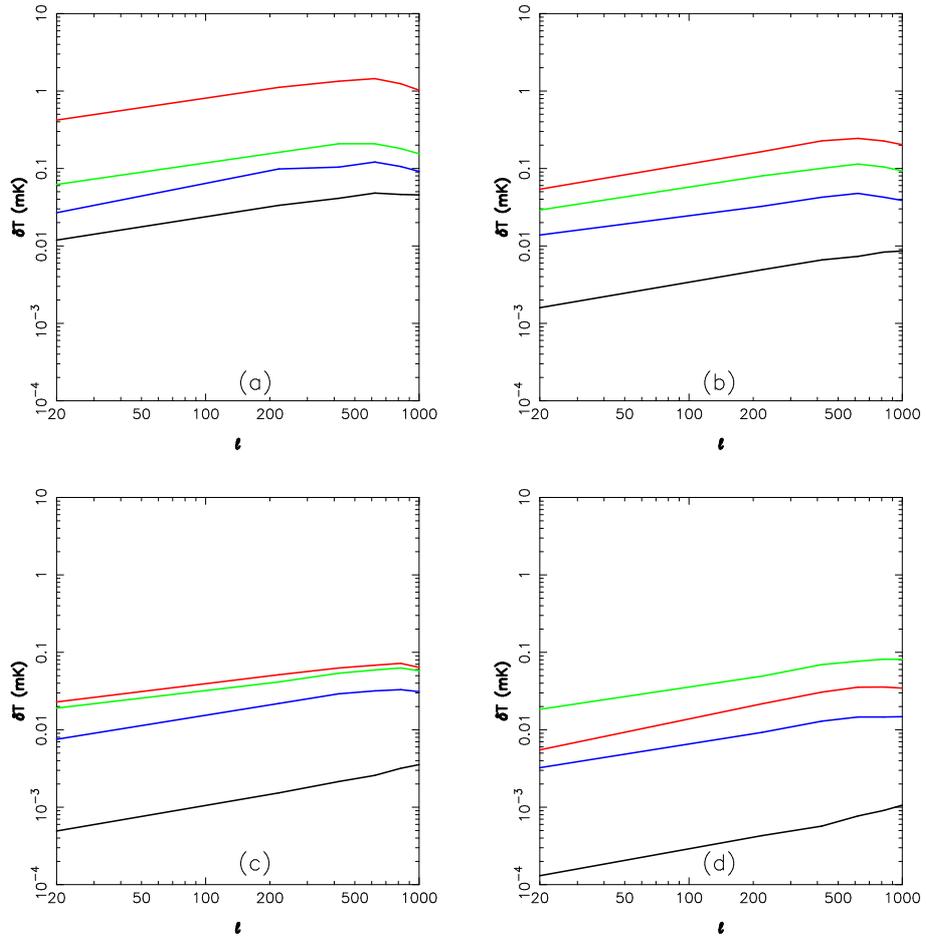}
\end{center}
\caption{The same as Fig. 7 but for the soft transition model of $w=200$ MHz.}
\end{figure}

\begin{figure}
\begin{center}
\psfig{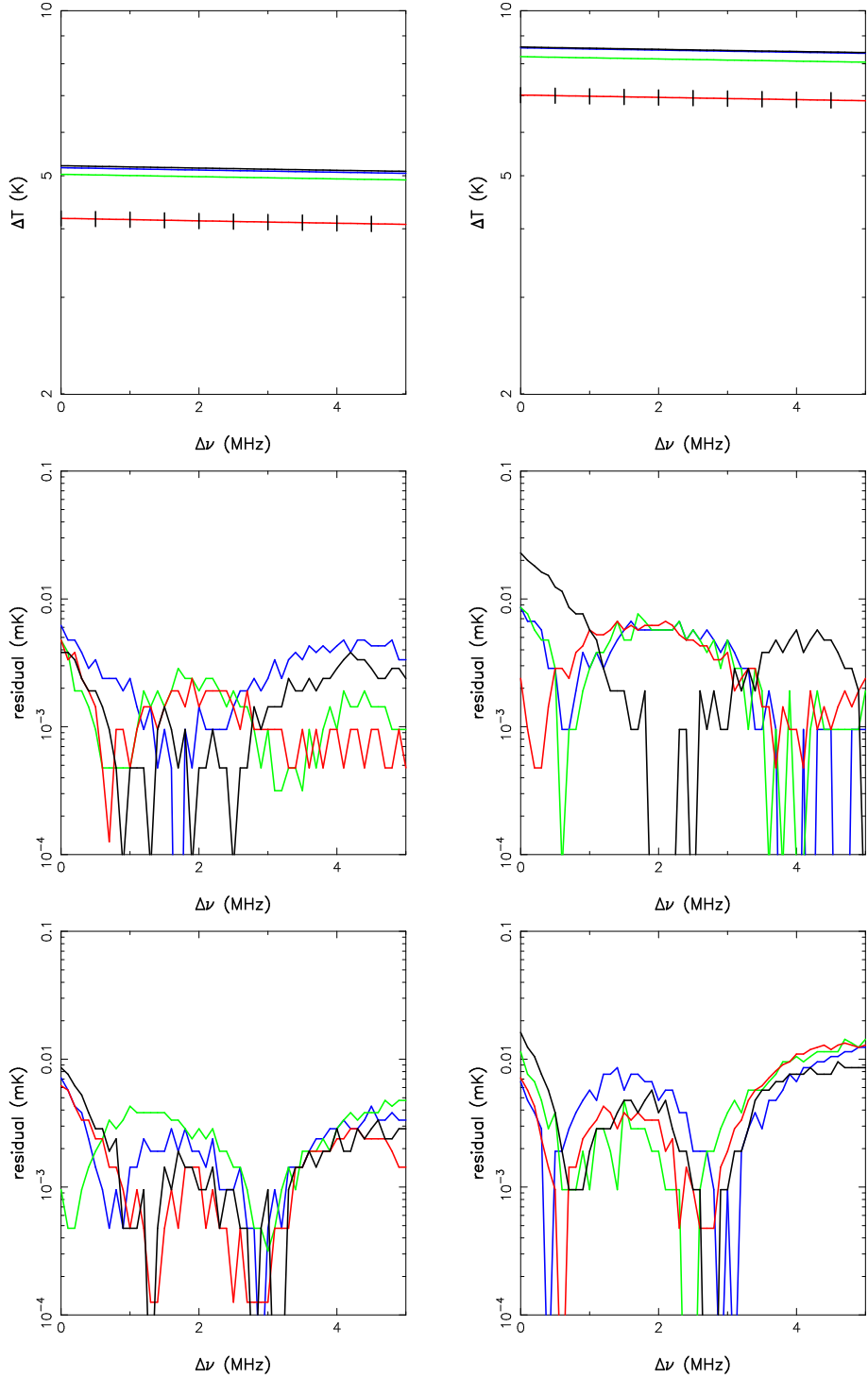}
\end{center}
\caption{Top panels: The multi-frequency angular power spectra 
against frequency interval $\Delta\nu$
for soft transition model, which are represented by 
$\Delta T=[\ell(2\ell+1)C_{\ell}(\Delta\nu)/4\pi]^{1/2}$; 
Middle panels: The residuals in $\Delta T$  after the best-fit
polynomials in the frequency intervals of 0-5 MHz are subtracted; 
Bottom panels: The residuals in $\Delta T$ after the best-fit 
polynomials in the frequency intervals of 0.5-5 MHz are subtracted. 
Left and right panels correpond to $\ell=400$ and $\ell=1200$, respectively.
The same colours as in Fig.4 are used to denote the different choices 
of $f$ parameters. We have illustrated, for the case of $f=10\%$,  
the typical statistical errors 
associated with the multi-frequency angular power spectra.}
\end{figure}

\end{document}